# A Hydrodynamic Interpretation of Quantum Mechanics via Turbulence

Roumen Tsekov[1], Eyal Heifetz[2] and Eliahu Cohen[3]

[1]Department of Physical Chemistry, University of Sofia, 1164 Sofia, Bulgaria
[2]Department of Geosciences, Tel-Aviv University, Tel-Aviv, Israel
[3]Faculty of Engineering and Institute of Nanotechnology and
Advanced Materials, Bar Ilan University, Ramat Gan, Israel

A stochastic Euler equation is proposed, describing the motion of a particle density, forced by the random action of virtual photons in vacuum. After time averaging, the Euler equation is reduced to the Reynolds equation, well studied in turbulent hydrodynamics. It is shown that the average pressure is nonlocal and the magnitude of the turbulent flow obeys the Fick law. Using the Madelung transformation, the Schrödinger equation is derived without any other assumptions.

Intensive studies are still devoted nowadays to the hydrodynamic interpretation of quantum mechanics, which originates from the seminal paper of Madelung,[1] published more than 90 years ago, and followed by Bohm and Vigier[2] who tried to introduce fluctuations in the Madelung fluid. The short analysis below attempts to continue this line of research, in the vicinity of turbulence arising from vacuum fluctuations.[3] Throughout this note, the vacuum is assumed to be relativistic and quantum, while the particle is assumed to be non-relativistic and classical.

Let us start first with a classical particle. According to classical mechanics, the particle's coordinate $\mathbf{R}(t)$ obeys the Newton law $m\ddot{\mathbf{R}} = -\partial_\mathbf{R} U$, where $U(\mathbf{R})$ is an external potential. Since we are looking for a hydrodynamic description, one can introduce the velocity $\mathbf{v} = \dot{\mathbf{R}}$ and local density $\rho(\mathbf{r}, t) = \delta(\mathbf{r} - \mathbf{R})$ of the particle. Via standard differentiations, the governing hydrodynamic mass and momentum balances are derived

$$\partial_t \rho + \nabla \cdot (\rho \mathbf{v}) = 0 \qquad \partial_t (m\rho \mathbf{v}) + \nabla \cdot (m\rho \mathbf{v}\mathbf{v}) = -\rho \nabla U \qquad (1)$$

being equivalent to Newton's second law. Quantum mechanics, however, is a probabilistic theory. Thus, the behavior of quantum particles is no longer deterministic and requires stochastic modeling. For eventually simulating a quantum behavior, we now augment Eq. (1) with a stochastic element.

According to modern physics, vacuum is filled with virtual particles mediating the physical interactions, which are present even for a free particle in a constant external potential. These virtual particles 'generate' the potential $U$ and behave stochastically as well.[4] Since vacuum is an isotropic medium, the virtual particles' fluctuations exert simply an external scalar pressure on the real particle. If the latter is a macroscopic one, the vacuum pressure fluctuations are assumed to cancel each other on the particle's surface. For this reason, there is no pressure term in Eq. (1). However, if the particle is very small, one arguably has to add heuristically the vacuum pressure fluctuations $p$ to obtain

$$\partial_t \rho + \nabla \cdot (\rho \mathbf{v}) = 0 \qquad \partial_t(m\rho \mathbf{v}) + \nabla \cdot (m\rho \mathbf{v}\mathbf{v}) = -\rho \nabla U - \nabla p \qquad (2)$$

Thus, the quantum particle no longer obeys the deterministic Newton law, and its random motion is governed by the stochastic Euler equation (2). Due to the lack of an average pressure in vacuum, the pressure fluctuations possess zero mean value, i.e. $\int p d^3 r \equiv 0$.

Further, it is expected that $p$ is a very rapidly fluctuating random pressure, which reflects directly into an irregular behavior of the hydrodynamic velocity $\mathbf{v}$ as well. Hence, the flow described by Eq. (2) is turbulent. This is also facilitated by the lack of viscous stress in Eq. (2) and by the high speed of the 'quantum' particle, due to its small mass $m$. Since the vacuum is stationary, one can apply the well-known time averaging procedure, developed for compressible fluids.[5] Employing the Reynolds temporal mean value $\overline{f}$ and Favre density weighted mean value $\tilde{f} \equiv \overline{\rho f}/\overline{\rho}$ the equations above transform after time-averaging to

$$\partial_t \overline{\rho} + \nabla \cdot (\overline{\rho} \tilde{\mathbf{v}}) = 0 \qquad \partial_t(m\overline{\rho}\tilde{\mathbf{v}}) + \nabla \cdot (m\overline{\rho}\tilde{\mathbf{v}}\tilde{\mathbf{v}} + \boldsymbol{\tau}) = -\overline{\rho}\nabla U - \nabla \overline{p} \qquad (3)$$

The well-known Reynolds stress $\boldsymbol{\tau} = \overline{m\rho \mathbf{v}\mathbf{v}} - m\overline{\rho}\tilde{\mathbf{v}}\tilde{\mathbf{v}} \equiv m\overline{\rho}\tilde{\mathbf{u}}\tilde{\mathbf{u}}$ defines the rms magnitude $\tilde{\mathbf{u}}$ of the turbulent velocity fluctuations. Hence, Eq. (3) reaches after rearrangements the form

$$\partial_t \overline{\rho} + \nabla \cdot (\overline{\rho} \tilde{\mathbf{v}}) = 0 \qquad \partial_t \tilde{\mathbf{v}} + \tilde{\mathbf{v}} \cdot \nabla \tilde{\mathbf{v}} = -\nabla U/m - \nabla \cdot (\overline{p}\mathbf{I} + m\overline{\rho}\tilde{\mathbf{u}}\tilde{\mathbf{u}})/m\overline{\rho} \qquad (4)$$

where $\mathbf{I}$ is the unit tensor. The ratio of the modulus of $\tilde{\mathbf{u}}$ and $\tilde{\mathbf{v}}$ gives the turbulence intensity.

Generally, there are four fundamental interactions in the Universe but the biggest success of quantum mechanics is in electrodynamics. Therefore, the most familiar force carriers seem to be virtual photons. Moreover, photons are the only free particles moving with the speed of light. The carriers of strong interaction (gluons) are never observed as free particles, while the gravitons have not been directly observed yet. The photons are stable and they do not interact with each other unless a creation/annihilation process is involved. According to quantum mechanics, the virtual photons cannot be removed from vacuum, since they carry the zero-point energy. The idea that the latter drives quantum mechanics is explored in the stochastic electrodynamics (SED) scheme.[6] Unless one introduces some cutoff, the virtual photons possess all frequencies $\omega$ and momenta $\hbar\omega/2c = \hbar k/2$ as well. Due to collisions with the virtual photons and momentum conservation, the real particle acquires the same momentum fluctuations. Hence, the speed of sound in the fluid is equal to $\hbar k/2m$. Note that it is the phase velocity, while the group velocity is $\hbar k/m$. The description above takes place in Fourier space, because the virtual photons are waves as well. Therefore, the Fourier image of the averaged pressure fluctuations, exerted by the virtual photons on the real particle, is given by $\bar{p}_k = m\bar{\rho}_k(\hbar k/2m)^2$. The latter looks like an ideal gas one, since quasi-particles like virtual photons are not interacting with each other in the Fourier space in the low energy regime. Applying the inverse Fourier transformation, the latter changes to[7]

$$\bar{p} = -(\hbar^2/4m)\nabla^2\bar{\rho} \qquad \int \bar{p}d^3r = 0 \qquad (5)$$

As easily seen, $\bar{p}$ is nonlocal, which could shed light on interference phenomena (and entanglement in multipartite scenarios), e.g. the double-slit experiment. As implied by Eq. (5), the pressure could be either positive or negative, which is due to the effect of the virtual photons in the ordinary space. For instance, if one considers a box with vacuum, there is a negative pressure trying to shrink the box, which is due to the zero-point energy via the Casimir effect. The mean pressure in vacuum, averaged along an infinite time as well, is zero since the zero-point energy $\hbar\omega/2$ tends to zero at $\omega \to 0$. The turbulent internal energy $\tilde{\varepsilon} = tr(\boldsymbol{\tau})/2\bar{\rho} = m\tilde{\mathbf{u}}^2/2$ is due to the external pressure $\bar{p}$. They are related via the first law of thermodynamics, $<\delta\tilde{\varepsilon}> = -<\bar{p}\delta\bar{\rho}^{-1}>$, where the brackets indicate statistical averaging $<f> = \int f\,\bar{\rho}d^3r$. Note that the heat is zero here, since the temperature of vacuum in zero as well. Substituting Eq. (5) in this equation yields an expression for the magnitude of the turbulent fluctuation velocity

$$\tilde{\mathbf{u}} = -(\hbar/2m)\nabla \ln \overline{\rho} \qquad\qquad \mathbf{u} = -(\hbar/2m)\nabla \ln \rho \qquad\qquad (6)$$

Remarkably, it describes a Fick flow $\overline{\rho}\tilde{\mathbf{u}} = -(\hbar/2m)\nabla\overline{\rho}$, where the universal Nelson[8] diffusion coefficient $\hbar/2m$ plays the role of a turbulent diffusivity. This is also the reason for the Heisenberg-like equality $<m\tilde{\mathbf{u}}\cdot\mathbf{r}> = 3\hbar/2$. From the derivation above one may deduce that the Planck constant $\hbar$ is a property solely of the virtual photons, while the particle is simply a tracer of the vacuum fluctuations. Alternatively, the pressure $\overline{p} = (\hbar/2)\nabla\cdot(\overline{\rho}\tilde{\mathbf{u}})$ is proportional to the divergence of the turbulent Fick flow. Thus Eqs. (4) can be written also in the form

$$\partial_t \overline{\rho} + \nabla\cdot(\overline{\rho}\tilde{\mathbf{v}}) = 0 \qquad\qquad \partial_t \tilde{\mathbf{v}} + \tilde{\mathbf{v}}\cdot\nabla\tilde{\mathbf{v}} = -\nabla U/m - \nabla\cdot(\hbar\overline{\rho}\nabla\tilde{\mathbf{u}}/2)/m\overline{\rho} \qquad (7)$$

suggesting that $\hbar\overline{\rho}/2$ is the turbulent dynamic viscosity, which is independent of the particle's mass. Note, that it multiplies the turbulent velocity gradient, not the hydrodynamic one $\nabla\tilde{\mathbf{v}}$ as in the standard closure for turbulent incompressible viscous flows.[5] The reason is, however, obvious, since quantum mechanics is a time reversible theory. Like in ideal gases, the turbulent kinematic viscosity coincides with the turbulent diffusivity $\hbar/2m$.

Let us consider now the turbulent energy balance

$$\partial_t(\overline{\rho}\tilde{\varepsilon}) + \nabla\cdot(\overline{\rho}\tilde{\varepsilon}\tilde{\mathbf{v}} + \mathbf{q} + \mathbf{C}) = -\overline{p}\nabla\cdot\tilde{\mathbf{v}} - \boldsymbol{\tau}:\nabla\tilde{\mathbf{v}} \qquad (8)$$

where $\mathbf{q} \equiv \overline{m\rho(\mathbf{v}-\tilde{\mathbf{v}})(\mathbf{v}-\tilde{\mathbf{v}})^2}/2$ and $\mathbf{C} \equiv \overline{p(\mathbf{v}-\tilde{\mathbf{v}})}$ are the turbulent heat flux and the pressure-velocity correlation, respectively. Since there are no energy dissipation in unitary quantum mechanics, the heat flux should be zero $\mathbf{q} = 0$, which resembles the symmetry of the velocity fluctuations in ideal fluids. Introducing the expressions for $\tilde{\varepsilon}$, $\overline{p}$ and $\boldsymbol{\tau}$ in Eq. (8) yields after rearrangements the pressure-velocity correlation

$$\mathbf{C} = -\overline{\rho}\tilde{\mathbf{u}}(\hbar\nabla\cdot\tilde{\mathbf{v}}/2) \qquad\qquad \int \mathbf{C}d^3r = -\frac{\hbar^2}{4m}<\nabla\nabla\cdot\tilde{\mathbf{v}}> \qquad (9)$$

As seen, it is an anti-correlation,[4] which shows that the average magnitude of the energy fluctuations, driving the pressure, is $-\hbar \nabla \cdot \tilde{\mathbf{v}}/2$. This is again the zero-point energy expression, where, according to the continuity equation (4), $-\nabla \cdot \tilde{\mathbf{v}} = d_t \ln \bar{\rho}$ is the typical frequency of the mean hydrodynamic flow and density fluctuations. The pressure-velocity correlation (9) transmits energy between the turbulent $\tilde{\mathbf{u}}$ and laminar $\tilde{\mathbf{v}}$ hydrodynamic velocities.

Finally, inserting Eq. (6) into Eq. (7) yields

$$\partial_t \bar{\rho} + \nabla \cdot (\bar{\rho}\tilde{\mathbf{v}}) = 0 \qquad \partial_t \tilde{\mathbf{v}} + \tilde{\mathbf{v}} \cdot \nabla \tilde{\mathbf{v}} = -\nabla(U+Q)/m \tag{10}$$

where $Q = \tilde{\varepsilon} + \bar{p}/\bar{\rho}$ is the fluid enthalpy, which is the thermodynamic potential for adiabatic systems at constant pressure. Indeed, the first law of thermodynamics expressed by the enthalpy reads $<\delta Q> = <\bar{\rho}^{-1}\delta\bar{p}> = 0$. Interestingly, $Q = -\hbar^2 \nabla^2 \bar{\rho}^{1/2} / 2m\bar{\rho}^{1/2}$ coincides with the well-known Bohm[9] quantum potential. Supposing also a potential hydrodynamic flow $\tilde{\mathbf{v}} = \nabla S/m$, Eqs. (10) can be further integrated to obtain

$$\partial_t \bar{\rho} + \nabla \cdot (\bar{\rho}\nabla S/m) = 0 \qquad \partial_t S + (\nabla S)^2/2m + U + Q = 0 \tag{11}$$

Introducing now the complex wave function via the Madelung[1] ansatz $\psi \equiv \bar{\rho}^{1/2} \exp(iS/\hbar)$, Eqs. (11) straightforwardly reduce to the Schrödinger equation

$$i\hbar \partial_t \psi = -\hbar^2 \nabla^2 \psi / 2m + U\psi \tag{12}$$

The quantum velocity operator, which is the inverse Fourier image of the group velocity $\hbar k/m$, generates a complex velocity $-i\hbar \nabla \psi / m = (\tilde{\mathbf{v}} + i\tilde{\mathbf{u}})\psi$ with the real and imaginary parts, being the hydrodynamics velocity and the turbulent magnitude, respectively. Thus, the density of the particle kinetic energy reads $m\bar{\rho}(\tilde{\mathbf{v}}^2 + \tilde{\mathbf{u}}^2)/2$. For this reason, an alternative derivation of quantum mechanics was suggested involving an imaginary stochastic process.[10] What remained unclear is the characteristic time of the pressure fluctuations, being the typical time constant $\tau$ in the Reynolds-Favre averaging procedure. Obviously, the average size of turbulent eddies is equal to $l = (\hbar \tau / m)^{1/2}$ due to the diffusive character of Eq. (6). On the other hand, this distance should be equal to the distance $l = c\tau$ traveled by the virtual photons. Solving these two equations together yields expressions for the time constant $\tau = \hbar/mc^2$ and the average size $l = \hbar/mc$ of eddies. Therefore, the pressure $p$ fluctuates with the Compton frequency and the typical eddy size is the reduced Compton wavelength.[11] Obviously, the size $d$ of a 'quantum' particle should be smaller than an eddy, i.e. $d < l$. Thus, one arrives to a mass-size criterion $md < \hbar/c$ for quantum behavior, where the right-hand side depends on properties of the virtual photons only. In general, Eq. (1) is valid for mass points. It is widely used in practice, however, to describe real particles, which possess finite size in any case. This approximation is justified if the typical length of the hydrodynamic gradients in the system is larger than the particle size.

To get a better perspective,[12] let us consider a Gaussian wave packet with normal distribution density $\bar{\rho} = \exp(-r^2/2\sigma^2)/(2\pi\sigma^2)^{3/2}$, whose dispersion $\sigma^2 = \sigma_0^2 + (\hbar t/2m\sigma_0)^2$ increases in time. The corresponding hydrodynamic $\tilde{\mathbf{v}} = \mathbf{r}\dot{\sigma}/\sigma$ and turbulent $\tilde{\mathbf{u}} = \mathbf{r}\hbar/2m\sigma^2$ velocities are collinear. Therefore, the turbulence intensity $\tilde{\mathbf{u}}/\tilde{\mathbf{v}} = 2m\sigma_0^2/\hbar t$ decreases in time and does not depend on the position. The pressure $\bar{p}/2\bar{\rho} = <\tilde{\varepsilon}> - \tilde{\varepsilon} = Q - <Q>$ is proportional to the turbulent kinetic energy fluctuations, which compensate exactly the enthalpy fluctuations of the particle. The pressure-velocity correlation is $\mathbf{C} = -2<\tilde{\varepsilon}>\bar{\rho}\tilde{\mathbf{v}}$. The second example is the well-known 1s orbital of a hydrogen atom $\psi(r) = \exp(-r/a_0)/(\pi a_0^3)^{1/2}$, where $a_0$ is the Bohr radius. The hydrodynamic velocity $\tilde{\mathbf{v}} = 0$ is zero, since $\psi$ is a real stationary solution. The magnitude of the turbulent velocity possesses only a radial component $\tilde{\mathbf{u}}_r = \hbar/ma_0$, which is constant everywhere. Thus, the turbulent internal energy $\tilde{\varepsilon} = \hbar^2/2ma_0^2$ is equal to the full kinetic energy of the electron. In the Bohr model, the electron rotates around the proton on a plane. According to the Schrödinger equation, the 1s electron is no longer rotating, which corresponds well to $\tilde{\mathbf{v}} = 0$. In Bohmian mechanics ($\dot{\mathbf{R}} \equiv \tilde{\mathbf{v}} = 0$), it stays fixed at $a_0$. Our explanation is that the electron moves randomly, thus drawing a picture equivalent to the spherical plot of the probability density. The pressure-velocity correlation is naturally zero. The ratio $l/a_0 = \alpha \approx 0.0073$ between the sizes of an eddy and of the 1s orbital is equal to the fine-structure constant. Interestingly, the ratio $r_e/l = \alpha$ between the electron radius and the eddy size is the same. Usually the transition to turbulence is modeled by the bifurcation theory, where the Feigenbaum constant, $\delta \approx 4.6692$, plays a central role. From this perspective it is exciting that $2\pi\alpha\delta^2 \approx 1.0000$. Perhaps bifurcations are the origin of the fundamental fine-structure constant. Since they are usually a scenario of turbulence growth, this could strengthen the turbulent hydrodynamic origin of quantum mechanics. This origin is somewhat supported by the intensive studies on classically-driven particles via hydrodynamics, which can mimic, to some extent, quantum motion.[13, 14]